%% file: quirks_tevatron.tex
\documentclass[aps,prl,twocolumn,showpacs,superscriptaddress,groupedaddress]{revtex4}  
\usepackage{graphicx}    % needed for figures
\usepackage{dcolumn}    % needed for some tables
\usepackage{bm}             % for math
\usepackage{amssymb}   % for math
\usepackage{multirow}

\begin{document}
\newcommand{\g}{{\gamma}}  
\newcommand{\Pt}{{p_T}}   
\newcommand{\Et}{{E_T}}   
\newcommand{\pT}{{p_T}}   
\newcommand{\e}{{\epsilon}}   
\newcommand{\MET}{$\not\!\!E_T$}
\newcommand{\dedx}{$\langle dE/dx \rangle$}
\newcommand{\met}{\not\!\!E_T}
\newcommand{\hmet}{\not\!\!H_T}

\hspace{5.2in} \mbox{FERMILAB-PUB-10-324-E}
\title{Search for new fermions (``quirks") at the Fermilab Tevatron Collider}

\input author_list.tex

\date{August 20, 2010}

\begin{abstract}
We report results of a search for particles with anomalously high ionization in events with
a high transverse energy jet and large missing 
transverse energy in $2.4$ fb$^{-1}$  of integrated luminosity collected by the D0 experiment at 
the Fermilab Tevatron $p\bar{p}$ collider. Production of such particles (quirks) is expected
in scenarios with extra QCD-like $SU(N)$ sectors, and this study is
the first dedicated search for such signatures.
We find no evidence of a signal and set a lower mass limit of 107~GeV, 119~GeV and 133~GeV for the mass of a charged quirk with strong dynamics scale $\Lambda$ in the range from 10 keV to 1 MeV and $N=2,3,$ and 5, respectively.
\end{abstract}

\pacs{14.80.Ly, 12.60.Jv, 13.85.Rm}
\maketitle 

One of the possible minimal extensions of the standard model (SM) is an addition of a new unbroken $SU(N)$ 
gauge group \cite{okun, bjorken} with new fermions, $Q$ and $\bar{Q}$, in the fundamental representation.
Such a SM extension has two parameters: the new fermion mass $m_Q$ and the strength of the new $SU(N)$ gauge 
coupling, infracolor, which can be defined through the scale $\Lambda$ where it becomes strong.

If the new fermions, quirks, carry SM charges, they can be pair-produced at colliders. The production cross-section will scale with $N$, the number of colors
in the new gauge group.  In principle, the quirks could transform
under any $SU(N)$ group, but $SU(2)$, $SU(3)$ and $SU(5)$ are considered
explicitly.
The phenomenology of such models at hadron colliders in the case $\Lambda \ll m_Q \simeq 0.1-1$ TeV has been recently studied 
in \cite{quirks}, revealing an array of exotic final states that could have eluded previous searches for physics beyond the SM. The defining feature is that breaking of the infracolor string is exponentially suppressed due to the large value of the ratio
$m_Q / \Lambda$. Unlike the SM quarks that immediately fragment into jets of hadrons, the quirk-antiquirk pair stays connected by the infracolor string like a rubber band that can stretch to macroscopic
length

\begin{equation}
L \simeq \frac{m_Q}{\Lambda^2} \simeq 1~\mu{\rm m} ~\left (\frac{m_Q}{100~{\rm GeV}}\right )\left (\frac{\Lambda}{100 ~{\rm keV}}\right)^{-2}.
\end{equation}

In this Letter, we consider a case where quirks have electric charge ($e$), no strong color charge, $10 ~{\rm keV} < \Lambda < 1~{\rm MeV}$, and $60~{\rm GeV} < m_Q < 1~{\rm TeV}$. This results in state with a decay length of many centimeters and a string size that is too small to be resolved in a detector, but large compared to atomic scales, e.g., for $m_Q=100$ GeV the corresponding string size is less than 100~$\mu$m, which is comparable to the resolution of the D0 tracking system. The quirk-antiquirk pair will be reconstructed in the detector as a single highly ionizing track. Since the net charge of the string is zero, the track will be straight. Its experimental signature will resemble that of a single very energetic muon.  However, due to low $\beta$, the pair will often reach the muon chamber outside of the muon trigger time window.  If it is produced in association with a high $\Et$ jet from initial state radiation, the event can be triggered and the track reconstructed. The signature which we consider is thus a high transverse momentum track with a large ionization-energy loss rate ($dE/dx$), a jet, and missing transverse energy ($\met$) aligned with the track. 

Several collider experiments have searched for highly ionizing slowly moving particles, and set limits on pair production of stable sleptons \cite{lep_gmsb}, stop quarks \cite{cdf_stop} and charginos \cite{lep_stable, fnal_champs}. Although some of the searches \cite{cdf_dedx, cdf_stop} were used to set an upper limit on the production cross section for a single, isolated, weakly interacting particle, the results can not be easily interpreted in the search for quirks.

The present search is performed on data collected between 2006 and 2008 with the D0 detector \cite{d0det} at the Fermilab Tevatron $p\bar{p}$ Collider at $\sqrt{s}=1.96$ TeV that correspond to 2.4 fb$^{-1}$ of integrated luminosity. A detailed description of the D0 detector can be found elsewhere \cite{d0det,smt, L0}. It comprises a central tracking system in a 2 T superconducting solenoid, a liquid-argon/uranium sampling calorimeter, and a muon spectrometer. The tracking system consists of a silicon microstrip tracker (SMT) and a central fiber tracker (CFT). The SMT, which extends from radius of $\simeq2$~cm to $\simeq10$~cm, has a six-barrel longitudinal structure,  each with a set of four layers arranged axially around the beam  and intersected with 16 radial disks. The silicon wafers are 300~$\mu$m thick with strip pitches $\simeq50~\mu$m and $\simeq150~\mu$m. In addition to position, the SMT detector provides a measurement of ionization energy  with a dynamic range up to $\simeq1.4$~MeV per strip ($\simeq 8$ times the energy deposited by a minimum ionizing particle moving perpendicularly to the silicon wafer).  The CFT, extending from a radius of $\simeq20$~cm to $\simeq50$~cm, has eight thin coaxial barrels, each supporting two doublets of overlapping scintillating fibers. The muon system,  extending from radius $\simeq3$~m to $\simeq6$~m,  resides beyond the calorimeter and consists of a layer of  tracking detectors and scintillating trigger counters before 1.8 T iron toroidal  magnets, followed by two similar layers after the toroids. The tracker readout gate is capable of detecting slowly moving particles with velocity $\beta\gtrsim0.1$ while in the muon system the readout gate is suitable only for particles with $\beta\gtrsim0.3$. The D0 detector uses a three-level trigger system to select events for the offline analysis. The events used in this analysis were recorded using a jet trigger requiring in addition a substantial missing transverse energy calculated using the sum of the jet momenta ($\hmet = |\sum_{\texttt{jets}}\vec{p}_T|$). Jets used in this analysis are reconstructed using the iterative midpoint cone algorithm \cite{jccb} with a cone size of 0.5. Jets must satisfy quality criteria which suppress background from leptons, photons, and detector noise effects.

A modified	 \cite{quirks} MadGraph~\cite{madgraph} event generator with {\sc pythia}
~\cite{pythia}  fragmentation and hadronization is used to simulate signal events, which are further processed with a {\sc geant}-based~\cite{geant} D0 detector simulation and  the same reconstruction as the data.  We investigate quirk masses ($m_Q$) in the range between 60 and 160 GeV and require that the jet from initial state radiation has a transverse momentum $p_T>65$ GeV. 
In the simulation, quirks are treated as non-interacting particles and their energy loss in the SMT is calculated outside of {\sc geant}. In the scenario considered in this study, the quirk-antiquirk pair forms a mesoscopic system. They follow a trajectory which is a classical superposition of the motion of the center of mass (c.o.m.) of the quirk-antiquirk system and the motion of the quirk (or the antiquirk) in the c.o.m. frame. The distribution of the velocities of the quirk-antiquirk system is very wide and peaks at $\beta\sim 0.8$ (0.2) for $m_Q=60$~(160)~GeV.  In order to estimate the specific energy loss $dE/dx$ of the quirk-antiquirk pair, we integrate the energy deposition of each quirk along its trajectory and normalize the sum of both contributions by the corresponding displacement of the c.o.m. The superposition of both particles and such nontrivial kinematics result in a large ionization ($\gtrsim15$~MeV/cm) with a distribution that is substantially wider than that of a lepton or a hadron. To simulate the effect of detector resolution and instrumental noise, we smear the calculated $dE/dx$ with the width of the $dE/dx$ distribution of muons from the $Z$ boson decays measured in data.  We simulate an effect of lost hits due to saturation that happens at $\Delta E = \frac{dE/dx\cdot l}{\cos{\theta}}\simeq\frac{1.4~{\rm MeV}}{\cos{\theta}}$, where $l=300~\mu$m is the thickness of a silicon wafer and where $\theta$ is the polar angle with respect to the proton beam direction and the quirk-antiquirk trajectory. 

The main SM backgrounds to the quirk signal are associated production of jets with a $W$ boson and multi-jet events. We select candidate events that have exactly one jet with $\Pt>75$ GeV and $|\eta|<1.6$~\cite{eta} and \mbox{ $\not\!\!E_T>50$}~GeV. Events with additional jets with $\Pt>25$~GeV are rejected to suppress multi-jet background.  We require a track in the fiducial region of the central tracker with $|\eta|<1.6$ and $\Pt>40$~GeV.  While the quirk-antiquirk
pair is expected to follow a straight trajectory, we allow for a non-zero reconstructed track curvature and require 
$p_T > 40$ GeV. The track should have at least four hits in the SMT and at least 12 hits in the CFT in order to suppress background from fake tracks.   Multi-jet events are further rejected by applying the following isolation criteria. The sum of the $\Pt$ of all other tracks in a cone of radius ${\cal R}=\sqrt{(\Delta\phi)^2+(\Delta\eta)^2}<0.5$ around the candidate track  must be less than 2.5~GeV. A similar isolation condition is applied for the total transverse energy measured in the calorimeter cells in a hollow cone of radius $0.1<{\cal R}<0.4$ around the candidate track; this energy must be less than 2.5~GeV. We require the track to be back-to-back in the transverse plane with the jet, $|\Delta\phi(trk, jet)|>2.5$.

\begin{figure}[hpb]
\begin{center}
\includegraphics[scale=0.4, angle=0]{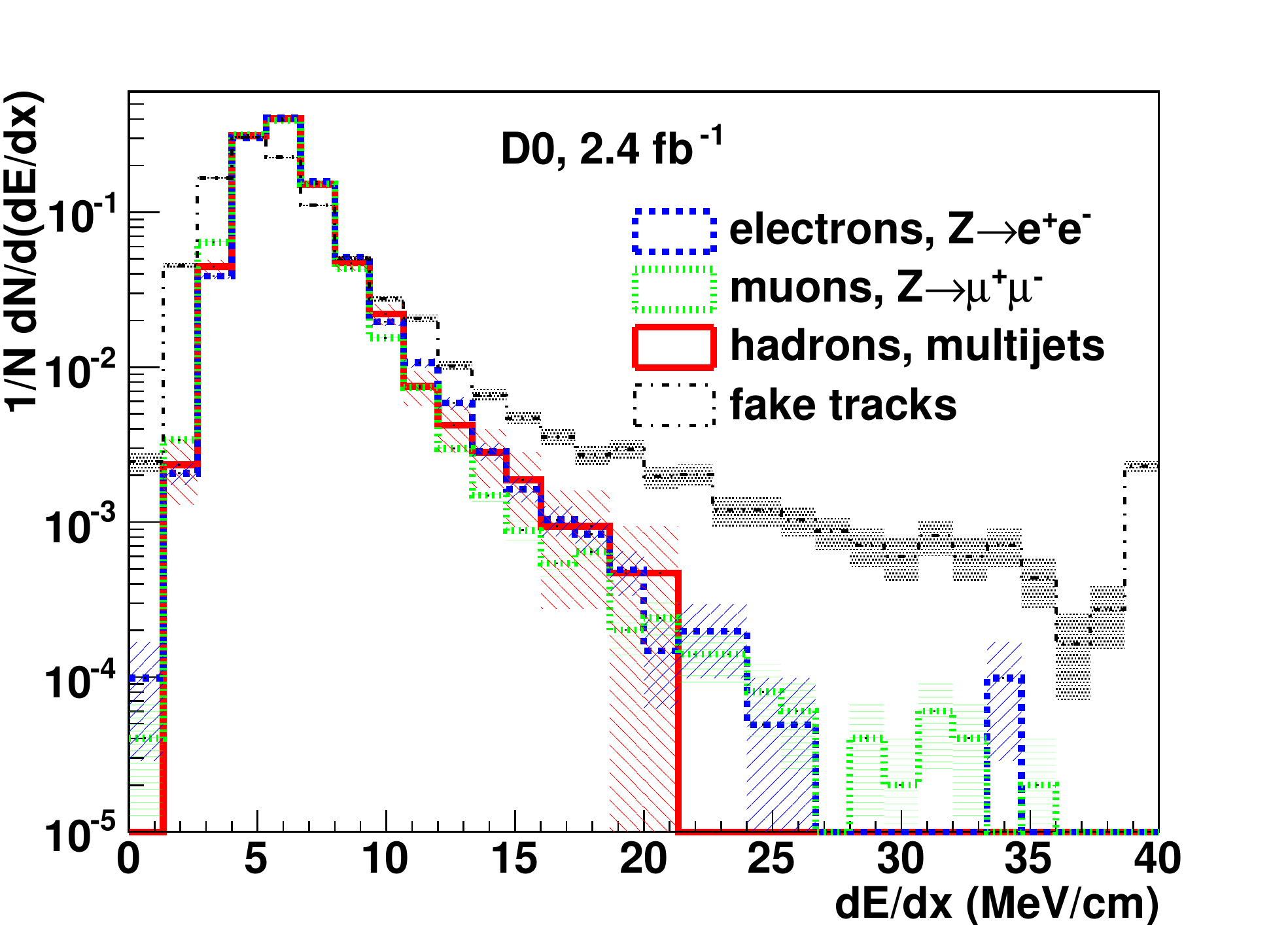}
\end{center}
\caption{(color online) Distribution of $dE/dx$ for electrons, muons, hadrons, and tracks made of random hits (fake tracks). Each distribution is normalized to unity and the last bin contains overflow events. The shaded area represents both the statistical and the systematic uncertainties.\label{fig:bkgs}}
\end{figure}

In order to suppress the $W(\rightarrow \ell\nu)+$jet background, we apply a cut on the azimuthal angle between the track and the missing transverse momentum, $|\Delta\phi~(trk,\not\!\!E_T)|<0.5$. We determine muon (electron) distributions of $dE/dx$ from $Z\rightarrow \mu^+\mu^-(e^+e^-)$ data events. In these events we require at least one of the leptons to have an associated track that passes all signal selection criteria and an invariant mass of the two leptons consistent with the $Z$ boson mass peak. The normalized distributions of $dE/dx$ along such tracks are shown in Fig.~\ref{fig:bkgs}.

The background events where a candidate track is produced by a charged hadron come from multi-jet events with incorrectly reconstructed $\not\!\!\!\!E_T$. We construct the $dE/dx$ distribution for these events from tracks found inside a jet (${\cal R} < 0.5$) with $\Pt > 40$~GeV and $|\eta|<1.6$. The track is required to pass the signal selection, including the track isolation requirement, and to originate from the same vertex as the jet ($|z_{jet}-z_{trk}|<1$~cm~\cite{eta}). 

Another background comes from fake tracks reconstructed from hit patterns due to combinatoric ambiguities. To determine the shape of the $dE/dx$ distribution, we use the same event selection criteria as those used for hadrons, except the requirement of zero CFT hits to suppress contribution from real charged particles (see Fig.~\ref{fig:bkgs}). The shape of the $dE/dx$ distribution for the fake tracks is significantly different from that for leptons and hadrons  since crossing angles for such tracks are mismeasured and thus the thickness per layer crossed is incorrectly estimated.

\begin{figure}[hpb]
\begin{center}
\includegraphics[scale=0.4, angle=0]{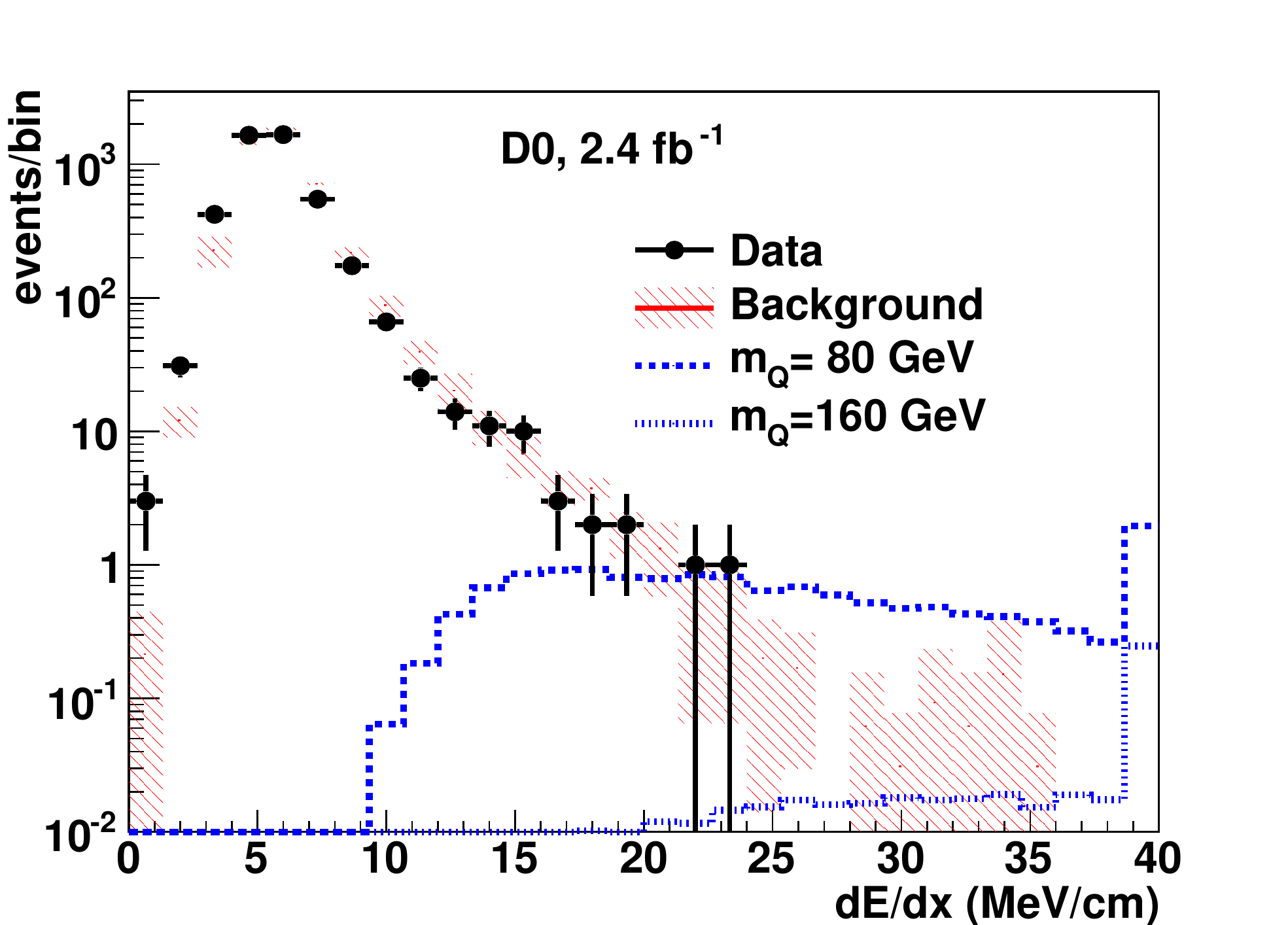}
\end{center}
\caption{(color online) Distribution of $dE/dx$ for candidate tracks in 4632 data events with overlaid background normalized to the number of events in data with $dE/dx<13$ MeV/cm. The shaded area represents the combined background uncertainty. The quirk signal is shown for two masses, $m_Q = 80~(160)$~GeV, and normalized to the expected number of events, 13 (0.53). The last bin contains overflow events. The shaded area represents both the statistical and the systematic uncertainties.\label{fig:final_dedx}}
\end{figure}

Fig.~\ref{fig:final_dedx} shows the comparison between the data and the average of only the lepton and hadron
backgrounds (without fake tracks), normalized to the region $dE/dx< 13$~MeV/cm, and the predicted quirk signal for two mass values. We observe no excess of highly ionizing tracks above the expected SM contribution and set limits on the quirk production cross-section.

We optimize the $dE/dx$ cut to achieve the lowest 95\%~C.L. expected limit for each quirk mass. The efficiency of the track quality requirement is obtained from $Z\rightarrow \mu^+\mu^-$ data events. Systematic uncertainties on the signal efficiency come from the combination of the uncertainties in modeling di-quirk $dE/dx$, trigger turn-on, and track quality, and are listed in Table~I. We take the largest variation of individual lepton and hadron backgrounds from their average as a systematic uncertainty on the background. 

We observe an excess of data at low $dE/dx$ with respect to the lepton and hadron-only background (see Fig~\ref{fig:final_dedx}). A fit of the data to a sum of fake track and SM background improves the agreement at low $dE/dx$. However, in the absence of signal we disregard the potential contribution from fake tracks to the events being considered as quirk candidates.

\begin{figure}[htp]
\begin{center}
\includegraphics[scale=0.4, angle=0]{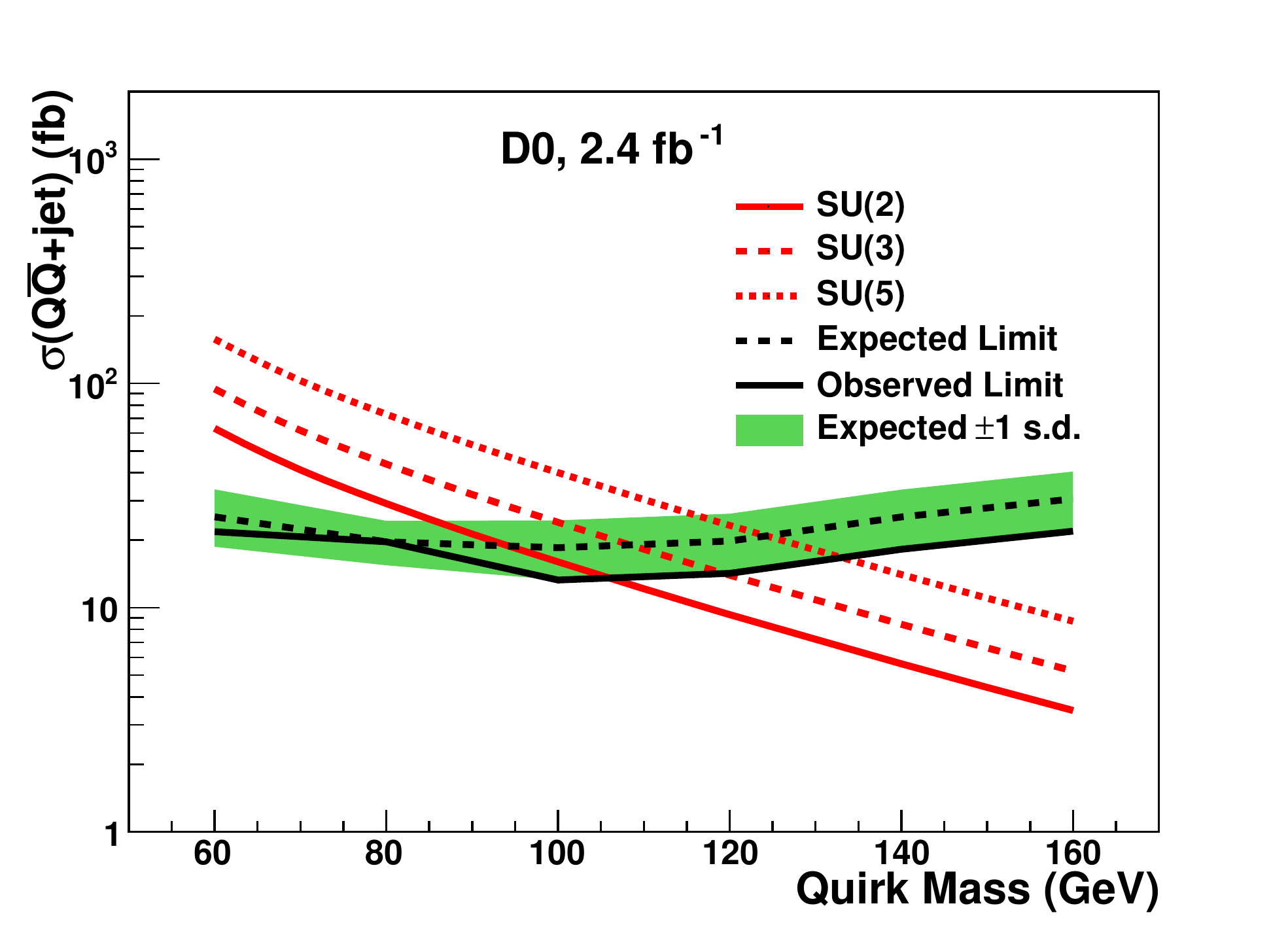}
\end{center}
\caption{(color online) Observed and expected 95\% C.L. limits on $\sigma(Q\bar{Q}+jet)$ for $SU(2)$, $SU(3)$ and $SU(5)$ gauge sectors. The band shows $\pm1$ standard deviation of the median expected limit. \label{fig:limits}}
\end{figure}
\input limits_table.tex

We perform a counting experiment in the region where $dE/dx$ is larger than a $m_Q$ dependent
threshold, given in Table I, and set an upper limit on the quirk production cross section at the 95\%~C.L. Limits are calculated using the modified frequentist  approach \cite{cls} and are consistent with expectation (Fig.~\ref{fig:limits} and Table~I). 

To summarize, we have performed a search for single highly ionizing tracks in events with an energetic jet and large missing transverse energy. This is the first study of this final state and the first search for quirk-antiquirk production. We find no excess of highly ionizing tracks and exclude charged quirks of mass up to 107~GeV, 119~GeV, and 133~GeV at the 95\% C.L. with strong dynamics scale $\Lambda$ in the range from 10 keV to 1 MeV for  $SU(2)$, $SU(3)$, and $SU(5)$ gauge sectors, respectively.  

\input acknowledgement.tex
\end{document}

%% file: author_list.tex
% remove these 3 lines before journal submittal.
%\centerline{author list dated 16 July 2010}
% end removal before journal submittal
%
\affiliation{Universidad de Buenos Aires, Buenos Aires, Argentina}
\affiliation{LAFEX, Centro Brasileiro de Pesquisas F{\'\i}sicas, Rio de Janeiro, Brazil}
\affiliation{Universidade do Estado do Rio de Janeiro, Rio de Janeiro, Brazil}
\affiliation{Universidade Federal do ABC, Santo Andr\'e, Brazil}
\affiliation{Instituto de F\'{\i}sica Te\'orica, Universidade Estadual Paulista, S\~ao Paulo, Brazil}
\affiliation{Simon Fraser University, Vancouver, British Columbia, and York University, Toronto, Ontario, Canada}
\affiliation{University of Science and Technology of China, Hefei, People's Republic of China}
\affiliation{Universidad de los Andes, Bogot\'{a}, Colombia}
\affiliation{Charles University, Faculty of Mathematics and Physics, Center for Particle Physics, Prague, Czech Republic}
\affiliation{Czech Technical University in Prague, Prague, Czech Republic}
\affiliation{Center for Particle Physics, Institute of Physics, Academy of Sciences of the Czech Republic, Prague, Czech Republic}
\affiliation{Universidad San Francisco de Quito, Quito, Ecuador}
\affiliation{LPC, Universit\'e Blaise Pascal, CNRS/IN2P3, Clermont, France}
\affiliation{LPSC, Universit\'e Joseph Fourier Grenoble 1, CNRS/IN2P3, Institut National Polytechnique de Grenoble, Grenoble, France}
\affiliation{CPPM, Aix-Marseille Universit\'e, CNRS/IN2P3, Marseille, France}
\affiliation{LAL, Universit\'e Paris-Sud, CNRS/IN2P3, Orsay, France}
\affiliation{LPNHE, Universit\'es Paris VI and VII, CNRS/IN2P3, Paris, France}
\affiliation{CEA, Irfu, SPP, Saclay, France}
\affiliation{IPHC, Universit\'e de Strasbourg, CNRS/IN2P3, Strasbourg, France}
\affiliation{IPNL, Universit\'e Lyon 1, CNRS/IN2P3, Villeurbanne, France and Universit\'e de Lyon, Lyon, France}
\affiliation{III. Physikalisches Institut A, RWTH Aachen University, Aachen, Germany}
\affiliation{Physikalisches Institut, Universit{\"a}t Freiburg, Freiburg, Germany}
\affiliation{II. Physikalisches Institut, Georg-August-Universit{\"a}t G\"ottingen, G\"ottingen, Germany}
\affiliation{Institut f{\"u}r Physik, Universit{\"a}t Mainz, Mainz, Germany}
\affiliation{Ludwig-Maximilians-Universit{\"a}t M{\"u}nchen, M{\"u}nchen, Germany}
\affiliation{Fachbereich Physik, Bergische  Universit{\"a}t Wuppertal, Wuppertal, Germany}
\affiliation{Panjab University, Chandigarh, India}
\affiliation{Delhi University, Delhi, India}
\affiliation{Tata Institute of Fundamental Research, Mumbai, India}
\affiliation{University College Dublin, Dublin, Ireland}
\affiliation{Korea Detector Laboratory, Korea University, Seoul, Korea}
\affiliation{CINVESTAV, Mexico City, Mexico}
\affiliation{FOM-Institute NIKHEF and University of Amsterdam/NIKHEF, Amsterdam, The Netherlands}
\affiliation{Radboud University Nijmegen/NIKHEF, Nijmegen, The Netherlands}
\affiliation{Joint Institute for Nuclear Research, Dubna, Russia}
\affiliation{Institute for Theoretical and Experimental Physics, Moscow, Russia}
\affiliation{Moscow State University, Moscow, Russia}
\affiliation{Institute for High Energy Physics, Protvino, Russia}
\affiliation{Petersburg Nuclear Physics Institute, St. Petersburg, Russia}
\affiliation{Stockholm University, Stockholm and Uppsala University, Uppsala, Sweden }
\affiliation{Lancaster University, Lancaster LA1 4YB, United Kingdom}
\affiliation{Imperial College London, London SW7 2AZ, United Kingdom}
\affiliation{The University of Manchester, Manchester M13 9PL, United Kingdom}
\affiliation{University of Arizona, Tucson, Arizona 85721, USA}
\affiliation{University of California Riverside, Riverside, California 92521, USA}
\affiliation{Florida State University, Tallahassee, Florida 32306, USA}
\affiliation{Fermi National Accelerator Laboratory, Batavia, Illinois 60510, USA}
\affiliation{University of Illinois at Chicago, Chicago, Illinois 60607, USA}
\affiliation{Northern Illinois University, DeKalb, Illinois 60115, USA}
\affiliation{Northwestern University, Evanston, Illinois 60208, USA}
\affiliation{Indiana University, Bloomington, Indiana 47405, USA}
\affiliation{Purdue University Calumet, Hammond, Indiana 46323, USA}
\affiliation{University of Notre Dame, Notre Dame, Indiana 46556, USA}
\affiliation{Iowa State University, Ames, Iowa 50011, USA}
\affiliation{University of Kansas, Lawrence, Kansas 66045, USA}
\affiliation{Kansas State University, Manhattan, Kansas 66506, USA}
\affiliation{Louisiana Tech University, Ruston, Louisiana 71272, USA}
\affiliation{University of Maryland, College Park, Maryland 20742, USA}
\affiliation{Boston University, Boston, Massachusetts 02215, USA}
\affiliation{Northeastern University, Boston, Massachusetts 02115, USA}
\affiliation{University of Michigan, Ann Arbor, Michigan 48109, USA}
\affiliation{Michigan State University, East Lansing, Michigan 48824, USA}
\affiliation{University of Mississippi, University, Mississippi 38677, USA}
\affiliation{University of Nebraska, Lincoln, Nebraska 68588, USA}
\affiliation{Rutgers University, Piscataway, New Jersey 08855, USA}
\affiliation{Princeton University, Princeton, New Jersey 08544, USA}
\affiliation{State University of New York, Buffalo, New York 14260, USA}
\affiliation{Columbia University, New York, New York 10027, USA}
\affiliation{University of Rochester, Rochester, New York 14627, USA}
\affiliation{State University of New York, Stony Brook, New York 11794, USA}
\affiliation{Brookhaven National Laboratory, Upton, New York 11973, USA}
\affiliation{Langston University, Langston, Oklahoma 73050, USA}
\affiliation{University of Oklahoma, Norman, Oklahoma 73019, USA}
\affiliation{Oklahoma State University, Stillwater, Oklahoma 74078, USA}
\affiliation{Brown University, Providence, Rhode Island 02912, USA}
\affiliation{University of Texas, Arlington, Texas 76019, USA}
\affiliation{Southern Methodist University, Dallas, Texas 75275, USA}
\affiliation{Rice University, Houston, Texas 77005, USA}
\affiliation{University of Virginia, Charlottesville, Virginia 22901, USA}
\affiliation{University of Washington, Seattle, Washington 98195, USA}
\author{V.M.~Abazov} \affiliation{Joint Institute for Nuclear Research, Dubna, Russia}
\author{B.~Abbott} \affiliation{University of Oklahoma, Norman, Oklahoma 73019, USA}
\author{M.~Abolins} \affiliation{Michigan State University, East Lansing, Michigan 48824, USA}
\author{B.S.~Acharya} \affiliation{Tata Institute of Fundamental Research, Mumbai, India}
\author{M.~Adams} \affiliation{University of Illinois at Chicago, Chicago, Illinois 60607, USA}
\author{T.~Adams} \affiliation{Florida State University, Tallahassee, Florida 32306, USA}
\author{G.D.~Alexeev} \affiliation{Joint Institute for Nuclear Research, Dubna, Russia}
\author{G.~Alkhazov} \affiliation{Petersburg Nuclear Physics Institute, St. Petersburg, Russia}
\author{A.~Alton$^{a}$} \affiliation{University of Michigan, Ann Arbor, Michigan 48109, USA}
\author{G.~Alverson} \affiliation{Northeastern University, Boston, Massachusetts 02115, USA}
% ...
%
% visitor_addresses.tex                        16 July 2010
%  available symbols are:
%  $\ast, \dag, \ddag, \S, \P, $\|$, $\ast\ast$, \dag\dag, \ddag\ddag ,\#
%
\collaboration{The D0 Collaboration\footnote{with visitors from
%{alton}
$^{a}$Augustana College, Sioux Falls, SD, USA,
%{burdin}
$^{b}$The University of Liverpool, Liverpool, UK,
%{evans,luty}
$^{c}$University of California Davis, Davis, CA,
%{haas,partridge}
$^{d}$SLAC, Menlo Park, CA, USA,
%{juste}
$^{e}$ICREA/IFAE, Barcelona, Spain,
%{luna-garcia}
$^{f}$Centro de Investigacion en Computacion - IPN, Mexico City, Mexico,
%{podesta-lerma}
$^{g}$ECFM, Universidad Autonoma de Sinaloa, Culiac\'an, Mexico,
and 
%{weber}
$^{h}$Universit{\"a}t Bern, Bern, Switzerland.%
%{hooper}
%$^{?}$%Visitor from Bradley University, Peoria, IL, USA.
%{kozminski
%$^{?}$}%Visitor from Lewis University, Romeoville, IL, USA.
%{deceased}
%$^{\ddag}$%Deceased.
}} \noaffiliation
\vskip 0.25cm

%% file: limits_table.tex
\begin{table*}
\begin{center}
\caption{Results for each simulated quirk mass: $dE/dx$ requirement, number of events in data, number of predicted background events,  overall signal efficiency, quirk plus jet production rate, and observed and expected 95\%~C.L. upper limits on the signal cross section.
\label{tab:summary}}
\begin{tabular}{c|c|c|c|c|ccc|c}
\hline
\hline

%&&&&&\multicolumn{3}{c|}{$\sigma_{SU(N)}$ (fb)}& \\

\multirow{2}{*} {$M_{\text{Q}}$ (GeV)} & 
\multicolumn{1}{c|}{${dE/dx}$ cut}& 
\multirow{2}{*} {$N_{\text{data}}$}&
\multirow{2}{*} {$N_{\text{bkg}}\pm\text{syst}$}&
\multirow{2}{*} {Efficiency$\pm\text{syst}$ (\%) }& 
\multicolumn{3}{c|}{ $\sigma_{SU(N)}$ (fb)} &
\multirow{2}{*} {Limit obs.(exp.) (fb)}
\\ 
  & (MeV/cm) & & & & $N=2$ & $N=3$ & $N=5$ & \\

\hline
%40  & 16 & 9  &$13\pm2$ & $ 9.3\pm1.9$ & 2986& 28 (41) \\ 
% \hline
60  & 19 & 4  &$5\pm1$ & $ 11\pm2.0$ & 63&94&157& 22 (25) \\ 
 \hline
80  & 21 & 2  &$1.9\pm0.8$ & $ 9.9\pm2.1$ & 29&43&72& 20 (20) \\ 
 \hline
100  & 24 & 0  &$0.9\pm0.4$ & $ 9.2\pm1.9$ & 16&24&60& 13 (18) \\ 
 \hline
120  & 24 & 0  &$0.9\pm0.4$ & $ 8.4\pm1.7$ & 9.3&14&23& 14 (19) \\ 
 \hline
140  & 24 & 0  &$0.9\pm0.4$ & $ 6.9\pm1.4$ & 5.6&8.4&14& 18 (25) \\ 
 \hline
160  & 24 & 0  &$0.9\pm0.4$ & $ 5.6\pm1.1$ & 3.4&5.1&8.5& 22 (31) \\ 
 \hline
\hline
\end{tabular}
\end{center}
\end{table*}

%% file: acknowledgement.tex
% acknowledgement.tex                             16 July 2010
%
We thank the staffs at Fermilab and collaborating institutions,
and acknowledge support from the
DOE and NSF (USA);
CEA and CNRS/IN2P3 (France);
FASI, Rosatom and RFBR (Russia);
CNPq, FAPERJ, FAPESP and FUNDUNESP (Brazil);
DAE and DST (India);
Colciencias (Colombia);
CONACyT (Mexico);
KRF and KOSEF (Korea);
CONICET and UBACyT (Argentina);
FOM (The Netherlands);
STFC and the Royal Society (United Kingdom);
MSMT and GACR (Czech Republic);
CRC Program and NSERC (Canada);
BMBF and DFG (Germany);
SFI (Ireland);
The Swedish Research Council (Sweden);
and
CAS and CNSF (China).

%% file: quirks_tevatron.bbl
\begin{thebibliography}{99}
\bibitem{eta}
Pseudorapidity $\eta$ is defined as $\eta =-\ln{\tan{\theta/2}}$, where $\theta$ is the polar angle with respect to the proton beam direction ($z$).
\end{thebibliography}
